\documentclass[a4paper,12pt,english]{article}

\usepackage{amsfonts,bm,amssymb,euscript,array,babel,cite}

 \usepackage[all]{xy}

\usepackage{amsmath,amsthm,amscd}

\usepackage[dvips]{epsfig}

\usepackage{slashed}

\usepackage[ansinew]{inputenc}



\newcommand{\be}{\begin{equation}}
\newcommand{\ee}{\end{equation}}

\newcommand{\beq}{\begin{equation}}

\newcommand{\eeq}{\end{equation}}

\newcommand{\beqa}{\begin{eqnarray}}

\newcommand{\eeqa}{\end{eqnarray}} 

\def\nn{\nonumber} \def \bea{\begin{eqnarray}} \def\eea{\end{eqnarray}}

\newcommand{\barr}{\begin{array}}

\newcommand{\earr}{\end{array}}

\numberwithin{equation}{section}

  \def\G{\Gamma}

 \def\d{\delta}

\def\mc{\mathcal}

\def\Z{{\mathbb Z}} \def\one{\mbox{1 \kern-.59em {\rm l}}}

\def\bit{\begin{itemize}} \def\eit{\end{itemize}}

\def\({\left(} \def\){\right)}

 \allowdisplaybreaks[3]

\begin{document}

\makeatother


\parindent=0cm

\renewcommand{\title}[1]{\vspace{10mm}\noindent{\Large{\bf

#1}}\vspace{8mm}} \newcommand{\authors}[1]{\noindent{\large

#1}\vspace{5mm}} \newcommand{\address}[1]{{\itshape #1\vspace{2mm}}}

\begin{titlepage}

\begin{center}

\title{ \Large Matrix theory origins of non-geometric fluxes}

\vskip 3mm

\authors{Athanasios {Chatzistavrakidis${}^{1}$} and Larisa {Jonke${}^{1,2}$}}

\vskip 3mm

\address{

{${}^1$}{\it Bethe Center for Theoretical Physics and Physikalisches Institut, University of Bonn, \\

Nussallee 12, D-53115 Bonn, Germany} 
 }

\address{{${}^2$}{\it Theoretical Physics Division, Rudjer Bo\v skovi\'c Institute, \\~Bijeni\v cka 54, 10000  Zagreb, Croatia} 
  }

\bigskip 

E-mails: than@th.physik.uni-bonn.de, larisa@irb.hr

\date{}

\vskip 1.4cm

\textbf{Abstract}

\vskip 3mm

\begin{minipage}{14cm}%

We explore the origins of  non-geometric fluxes within the context of M theory described 
as a matrix model. Building upon compactifications of Matrix theory on non-commutative tori and 
twisted tori, we formulate the conditions which describe compactifications with non-geometric fluxes. 
These turn out to be related to certain deformations of tori with non-commutative and 
non-associative structures on their phase space. Quantization of flux appears as a natural 
consequence of the framework and leads to the resolution of non-associativity at the level of the 
unitary operators.
The quantum-mechanical nature of the model bestows an important role on the phase space. 
In particular, the geometric and non-geometric fluxes exchange their properties when going from  
position space to momentum space thus providing a duality among the two. Moreover, 
the operations which connect solutions with different fluxes are described and their 
relation to T-duality is discussed. Finally, we provide some insights on the effective 
gauge theories obtained from these matrix compactifications.

\end{minipage}

\end{center}

\end{titlepage}

\tableofcontents

\section{Introduction}

Superstring theories offer an attractive framework for the ultraviolet completion of our current knowledge 
of nature as described by the standard model of particle physics and general relativity for gravitational 
interactions. As such, there is hope that superstring theories will ultimately account for
 physics both at the weak scale and at the Planck scale. A deeper conceptual unification of superstring 
theories, including the several dualities among themselves, is achieved in the context of M theory, whose
full quantum-mechanical incarnation remains however elusive. A very interesting proposal for its non-perturbative 
definition was provided in Ref. \cite{Banks:1996vh} and is known under the name of Matrix theory.

The attempt to connect superstring theories to our low-energy, four-dimensional world traditionally involves 
a compactification of the ten-dimensional theory and a subsequent dimensional reduction to four dimensions. 
Clearly, finding the correct vacuum which would reproduce the standard model at low energies is not an easy 
task. A lot of attention in the recent years focused on flux compactifications \cite{Grana,Kachru}, where the internal components 
of p-form fields, present in string theory, acquire a vacuum expectation value. This long-standing programme 
is mainly carried out in the supergravity approximation, which is the low-energy, field theory limit of 
perturbative string theory.

A separate development, mainly in the direction of providing a non-perturbative definition of superstring theories,
arose in the context of reduced matrix models \cite{EK}. In this framework, the dynamical variables are represented by 
large Hermitian matrices which provide the microscopical degrees of freedom of superstrings. In the case of the type 
IIB superstring theory such a model was proposed in Ref. \cite{Ishibashi:1996xs} and it was studied further in 
numerous instances. These range from the study of non-commutative Yang-Mills theories \cite{Aoki1}, 
the structure and dimensionality of spacetime \cite{Aoki2,Nishimura1,Kawai} and the emergence of 
geometry and gravity \cite{steinacker} to applications in particle physics \cite{Aoki3,Chatzistavrakidisinter} 
and cosmology \cite{Kim1,Kim2}. It is worth mentioning that these studies use both analytical tools and/or 
Monte Carlo simulations. A recent review of the latter with a more complete list of references is 
Ref. \cite{Nishimurarev}. 
Moreover, as already mentioned, the authors of 
Ref. \cite{Banks:1996vh} suggested a matrix model serving as a non-perturbative formulation of M theory, 
called Matrix theory.

In connection to the string compactification programme, a natural development was the study of 
compactifications in the framework of Matrix theory.
 The first systematic study was performed by Connes, Douglas and Schwarz in Ref. \cite{cds}, 
where toroidal compactifications of Matrix theory were examined and important relations to 
non-commutative geometry and non-commutative gauge theories were described. It was argued that 
matrix compactifications on non-commutative tori can be related to supergravity compactifications. 
In particular, a constant deformation of the torus leads to a theory which is tantamount to a vacuum of 
eleven-dimensional supergravity with constant background three-form potential. In type IIA language, 
there is a reciprocal relation among the constant non-commutativity parameter $\theta$ and a constant 
B-field.

The relation between non-commutativity parameters and background values of fields raises the question 
whether this analogy can be extended to more general situations, i.e. how flux compactifications can be understood 
in matrix models. Conventional string compactifications may include geometric fluxes and NS-NS fluxes, as well 
as R-R fluxes in the type II cases \cite{Grana,Kachru}. A first description of geometric fluxes in Matrix theory was given in 
Ref. \cite{ramgoolam} for the case of the three-dimensional twisted torus. This was recently 
revisited and generalized to 
 higher-dimensional twisted tori, utilizing their construction as quotients of nilpotent Lie groups by 
certain discrete subgroups of them (nilmanifolds) \cite{Chatzistavrakidis:2012yp}. However, the possibility 
of describing NS-NS flux compactifications in this framework has not been studied yet\footnote{See 
however Ref. \cite{Anazawa}.}. What is more, a lot 
of attention was drawn recently to the so-called non-geometric fluxes corresponding to unconventional 
compactifications whose origin is not yet fully understood. Clearly, their possible role in Matrix theory 
was not yet addressed.

Non-geometry is intimately connected to T-duality \cite{ng1,ng2,ng3}\footnote{For T-duality in Matrix theory, 
see for example the 
Refs. \cite{Ganor,bmz,bmz2}.}. Moreover, it may be related to an 
unconventional type of fluxes which can be present in the effective superpotential of a 
string compactification \cite{ngf1,ngf2}.  
A convenient way to think of such backgrounds
 has as starting point a 
toroidal compactification of string theory on a standard torus penetrated by a NS-NS flux $H_{ijk}$. 
As usual, let us restrict our discussion to a three-dimensional torus, keeping in mind that this is 
not a fully consistent background of string theory and has to be appropriately extended, as 
discussed in Ref. \cite{kstt}. It serves as a toy model, whose central properties may be directly transferred to a 
full-fledged vacuum.
Performing a T-duality along one direction of the torus the T-dual geometry is described by the three-dimensional 
nilmanifold, the twisted torus, whose non-trivial spin connection serves as a geometric flux. Such backgrounds 
lie in the heart of Scherk-Schwarz compactifications \cite{Scherk} and they were studied systematically in 
\cite{km} and more recently in \cite{kstt,Hulltts,Hullttm,prezasg2,Grana2}. A second T-duality along another direction of the 
torus takes the twisted torus background to a situation which is globally ill-defined. The fields of the 
theory cannot be patched with the usual transition functions anymore;
 instead this patching requires T-duality elements \cite{Hull1}. The situation gets even worse when a third T-duality 
is considered. This bizarre situation was tackled in the context of the doubled formalism,
where an extended space is considered such that duality transition functions become diffeomorphisms 
of the enlarged manifold \cite{Hull1}. Further progress led to the construction of twisted doubled tori, 
which provided a context where both geometric and non-geometric situations can be described
 \cite{hulltdt,prezastdt}. The above connection between NS-NS flux ($H$), geometric flux ($f$) and non-geometric 
fluxes ($Q$ and $R$ respectively) may be described by the following T-duality chain:
\be 
H_{ijk} \overset{T_k}\longrightarrow f_{ij}^{\ \ k} \overset{T_j}\longrightarrow Q_i^{\  jk} 
\overset{T_i}\longrightarrow R^{ijk}.
\label{chain}\ee
More recently, a ten-dimensional description of non-geometric fluxes was investigated in the context 
of generalized geometry \cite{Andriot1,Andriot2}, double field theory \cite{DF1,DF2,DF3,DF4} and non-associative geometry \cite{Blumenhagen1,Blumenhagen2}\footnote{After the first version of this paper was posted, Ref.\cite{bermannew} appeared, where non-geometric fluxes in M theory are discussed as well within the framework of  (M-theory extended)  generalized geometry.}.

From the early studies of non-geometric backgrounds it was realized that $Q$ fluxes are somehow associated to 
non-commutativity  and $R$ fluxes to non-associativity of the underlying space
 \cite{ngncna1,ngncna2,ngncna3,Grange}.
 Recently, 
the emergence of non-commutative and non-associative geometries in compactifications with non-geometric 
fluxes was described by L\"{u}st from a physically motivated perspective \cite{Lust1,Lust2}. 
In particular, it can be related to the properties of a quantum-mechanical particle 
moving under the influence of a (non-constant) magnetic field \cite{Jackiw1,Jackiw2}.
 This is also reminiscent of the Landau 
problem in quantum mechanics (see for example the discussion in Ref. \cite{Szaboreview}). Furthermore, the quantization 
of such backgrounds was recently elaborated in Ref. \cite{szabonong}. 
Such structures were also derived using conformal field theory in Ref.\cite{BP} and later also appeared in the context of asymmetric orbifolds \cite{CFL}.

Having already mentioned the 
close connection between non-commutativity and compactifications of Matrix theory, it is worth examining 
whether non-geometric fluxes can be traced in non-commutative and/or non-associative deformations of tori
in this context. This is the main topic of the present paper. 
We believe that there are some advantages in this programme as compared to 
investigations of such structures directly in supergravity. First of all, Matrix theory is inherently 
quantum-mechanical and phase space plays an important role in the study of its compactifications. Secondly, 
 supergravity, being a field theory, does not include the stringy winding modes. This is important because 
in non-geometric string backgrounds  momentum modes and winding modes appear to be mixed. Such aspects 
can be addressed in the non-perturbative context of Matrix theory \cite{Ho,Taylor}. Finally, another 
advantage of Matrix theory over supergravity regards the quantization of flux. Indeed, while in string 
theory charges are quantized, in supergravity, being a classical field theory, the charges are 
continuous parameters \cite{Grana}. On the contrary, in the context of Matrix theory we will determine appropriate 
quantization conditions.

The structure of the paper is as follows. In section 2 the matrix model of Banks, Fischler, Shenker and 
Susskind (BFSS) is briefly reviewed, along with its known compactifications on non-commutative tori 
and twisted tori. The issue of the description of non-geometric situations in Matrix theory is addressed 
in section 3. First, the algebraic building blocks responsible for the non-commutative/non-associative 
deformations are described by implementing the properties of twisted doubled tori in a matrix model 
framework. The way that each algebraic block can be obtained from another one indicates how T-duality 
operates on the corresponding solutions. In the process of relating such solutions we find a correspondence 
between position and momentum space which is reminiscent of a frame choice in generalized geometry. 
Furthermore, we obtain flux quantization conditions, a property 
that resolves the non-associativity of  unitary operators.
In section 4 certain aspects of the resulting gauge theories obtained from the compactifications of Matrix theory 
are discussed. 
 Finally, we summarize our findings in section 5.

\section{The matrix model and its compactifications}

\subsection{The BFSS matrix model}

Let us begin by briefly describing the 
BFSS matrix model \cite{Banks:1996vh}.
This model, also referred to as Matrix theory, was suggested as a non-perturbative definition of M theory. Its action, determining the dynamics of $N$ D0 branes in uncompactified 
spacetime, is given by the following functional:
\be\label{BFSSaction}
{\cal S}_{BFSS}=\frac 1{2g}\int dt\biggl[ Tr\big(\dot  {\mathcal X}_a\dot {\mc X}_a
-\frac 12 [{\mc X}_a,{\mc X}_b]^2\big)
+2\psi^T\dot\psi-2\psi^T\Gamma^a[\psi,{\mc X}_a]\biggl], 
\ee  
where $ \mc X_a(t),a=1,\dots,9$ are nine time-dependent $N\times N$ Hermitian matrices, $\psi$ are their fermionic superpartners 
and $\G^a$ furnish a representation of $SO(9)$. In the limit of infinite-dimensional matrices, 
namely $N\rightarrow \infty$, this matrix model is supposed to be equivalent to uncompactified M theory. 
In the following we shall be concerned mainly with the 
bosonic part of the above action. An important point is that this model may be thought of as supersymmetric 
quantum mechanics. As such it is inherently quantum-mechanical, which will play an important role in 
our discussions in the following sections.

The equations of motion resulting from the variation of the action (\ref{BFSSaction}) with respect to 
$ \mc X_a$, setting $\psi=0$, are 
\be \label{BFSSeom}
\ddot {\mc X}_a+[{\mc X}_b,[{\mc X}^b,{\mc X}_a]]=0,
\ee
where indices are raised and lowered with $\delta_{ab}$ and therefore it does not make any difference 
whether they are upper or lower. For static configurations it is clear that the first term in (\ref{BFSSeom}) 
may be dropped.

\subsection{Compactification on tori}

A matrix compactification on a $d$-dimensional torus is defined by 
a restriction of the matrix action under certain periodicity conditions incorporating the cycles of the 
torus. Let us restrict to the 3-dimensional case, since most of our following discussions will relate 
to this number of dimensions. The generalization to any dimension is simple and straightforward.

For a T$^3$ extending, say, in the directions $\mc X_1,\mc X_2,\mc X_3$, the compactification involves three invertible 
unitary matrices $U^i$ acting as translation operators and leading to the conditions
\bea \label{conditionst3}
\mc X_i+R_i&=&U^i\mc X_i(U^i)^{-1}, \quad i=1,2,3, \nn\\
\mc X_a&=&U^i\mc X_a(U^i)^{-1}, \quad a\ne i, \quad a=1,\dots,9, 
\eea where $R_i$ are constants. In the ensuing we set these constants to one since they may be easily 
reinserted at any point of the analysis.

It is well-known \cite{cds} that the conditions (\ref{conditionst3}) are solved by 
\bea \label{slnnctd}
\mc X_i&=&i{\mathcal D}_i,  \quad\mc X_m=\mathcal A_m(\hat U^i), \quad m=4,\dots,9, \nn\\
U^i&=&e^{i\hat x^i},
\eea
where the unitary operators satisfy 
$U^iU^j=\lambda^{ij}U^jU^i$
with complex constants  $\lambda^{ij}=e^{ -i\theta^{ij}}$, while the covariant derivatives have the form
\be \label{covder}
\hat{\mathcal D}_i=\hat\partial_i-i\mathcal A_i(\hat U^j).
\ee
 The special case of $\theta^{ij}=0$ or 
equivalently $\lambda^{ij}=1$ leads to commuting $U$s, which correspond to the
case of a standard $3$-torus T$^3$. 
However, in general the parameters $\theta^{ij}$ are not vanishing, in which case the $U$s are not commuting operators. 
The latter case corresponds to a compactification on a non-commutative torus.

Let us note that $\mc A_i$ and $\mc A_m$ in 
(\ref{slnnctd}) do not depend on $U$s, but rather on the set of operators $\hat U^i$ which commute with all $U^i$, i.e.
$
[\hat U^i, U^j] =0.
$ 
 This has to be true in order for the conditions (\ref{conditionst3}) to be satisfied. Of course, in the 
commutative case the $\hat U$ and the $U$ are identified.
Moreover, a direct implication of the non-commutativity among the $U^i$
is that the $\hat x^i$ do not commute as well, but instead 
they satisfy the relation
\be \label{3nc}
[\hat x^i, \hat x^j]= i\theta^{ij}.
\ee
Thus they may be interpreted as the coordinate operators of a non-commutative $3$-torus T$^3_{\theta}$, 
as long as they are periodic.
In addition, in the present case the $\hat U$s may be written as
\be 
\hat U^i=e^{i\hat x^i+\theta^{ij}\hat\partial_j},
\ee 
and they satisfy the relations
\be \label{cteht}
\hat U^i\hat U^j=e^{ -i \hat\theta^{ij}}\hat U^j\hat U^i, \quad \hat\theta^{ij}=-\theta^{ij}. \ee
Therefore, the compactification on a non-commutative torus leads to a gauge theory  on a dual non-commutative torus with parameter $\hat\theta$. Indeed, substituting the solution back to the original matrix model 
action, one obtains a non-commutative supersymmetric Yang-Mills theory on this dual torus \cite{bmz}. 

Let us stress that the above considerations involve the full phase space of $\hat x^i$ and 
$\hat p_i=-i\hat\partial_i$ and that the phase space algebra reads as
\bea 
[\hat x^i, \hat x^j] &=& i\theta^{ij}, \nn \\
{[}\hat x^i,\hat p_j] &=& i\delta^i_j, \nn\\
{[}\hat p_i, \hat p_j] &=& 0.
\label{psat}\eea
Note that the momenta are commutative and they satisfy the standard Heisenberg relation with the coordinates,
while the latter exhibit constant non-commutativity.

The constant non-commutativity among the coordinate operators, which also controls the non-commutativity 
properties of the unitary operators $U^i$ and $\hat U^i$, has an interesting physical interpretation. 
It corresponds to turning on a background value for the 3-form potential $C^{(3)}$ of 11-dimensional supergravity, 
which is the low-energy, field theory limit of M theory. Indeed, Connes, 
Douglas and Schwarz (CDS) suggested that that the deformation parameters $\theta^{ij}$ defining the non-commutative 
tori, correspond to moduli of the 11-dimensional supergravity, such that
\be \label{cdsm}
(\theta^{-1})_{ij}\propto \int dx^idx^j C^{(3)}_{ij-},
\ee
where ``$-$'' denotes the light cone direction $x^-$ \cite{cds}. In the language of the type IIA
theory, which is obtained from 11-dimensional supergravity upon compactification on a circle, this relation may
be written as  
\be \label{cds}
(\theta^{-1})_{ij}\propto \int dx^idx^j B_{ij},
\ee 
where the NS-NS 2-form field $B$ of the type IIA supergravity is obtained by the 3-form $C^{(3)}$ in the 
compactification process. We will often denote such relations as 
\be \label{cdssch}
\theta^{ij} \quad \overset{CDS}\longleftrightarrow \quad  B_{ij}~,
\ee 
where the left hand side is related to the BFSS matrix model and the right hand side to type II backgrounds.

\subsection{Compactification on twisted tori}

Following
the same reasoning as before, it is straightforward to define compactifications of Matrix theory 
on twisted tori. 
The simplest example of a twisted torus arises for $d=3$. In that case, a (twisted) compactification is achieved by imposing 
and solving an appropriately extended set of constraints, which involve again three unitary matrices $U^1,U^2,U^3$ and they are\footnote{Note that these constraints are slightly different than 
the ones presented in \cite{Chatzistavrakidis:2012yp}. This is nothing but an equivalent description 
of the 3-dimensional twisted torus which will render our present discussion more practical.}
\bea \label{conditionstt3}
U^i\mc X_i(U^i)^{-1}&=&\mc X_i+1, \quad i=1,2,3,\nn\\
U^1\mc X_3(U^1)^{-1}&=&\mc X_3-N\mc X_2 , \nn\\
U^2\mc X_3(U^2)^{-1}&=&\mc X_3+N\mc X_1, \nn\\
U^i\mc X_a(U^i)^{-1}&=&\mc X_a, \quad a\ne i, \quad a=1,\dots,9,  \quad (a,i)\ne \{(3,1),(3,2)\}.
\eea
The latter constraints generalize the ones for the square torus appearing in (\ref{conditionst3}), thus 
incorporating the twist of the 3-dimensional nilmanifold $\tilde {\text{T}}^3$ 
\cite{ramgoolam,Chatzistavrakidis:2012yp}. 
A general solution of the above constraints is  given again by Eq. (\ref{slnnctd}), with the following 
new features:
\begin{itemize}
 \item The operators $U^i$ satisfy the relation 
\bea\label{tt3ub}
U^iU^j&=&e^{- i\theta^{ij}-iNf_{\ k}^{ij}\hat x^k}U^jU^i,\eea
where $f^{ij}_{\ \ k}=f_k^{\ ij}$ are antisymmetric only in the upper two indices and 
the only non-vanishing components are $f^{12}_{\ \  3}=-f^{21}_{\ \ 3}=1$. In fact 
they correspond to the structure constants of the unique nilpotent Lie algebra in three dimensions, 
which plays an important role in the construction of the twisted torus \cite{Chatzistavrakidis:2012yp}. Moreover, these parameters are also known as geometric fluxes in the language of Scherk-Schwarz
 compactifications \cite{km}. The relation 
(\ref{tt3ub}) defines a non-commutative twisted torus. 
\item The components $\mc A_i$ must be corrected to 
${\mc{\hat A}}_i=\mc A_i+iNf_{i}^{\ jk}\mc A_j\hat\partial_k$, which in the present case means that only 
$\mc A_3$ has to be corrected accordingly.
\item For the above solution we can find  the set of $\hat U$s which provide the 
dependence of $\mc A_i$ and $\mc A_m$ and thus give the 
connection on a trivial gauge bundle. They have the form $\hat U^i=e^{i\hat y^i}$, where
\bea \label{gentwyy}
\hat y^i = \hat x^i-i\theta^{ij}\hat\partial_j-iNf^{ij}_{\ \ k}\hat x^k\hat\partial_j.
\eea
As before, these operators commute with the $U^i$ by construction\footnote{An important note on notation is in order. The operators $U^i$ and $\hat U^i$ act on a Hilbert space of states $\mc H$. Let an 
arbitrary element of $\mc H$ be $f(\hat x^i)$. Then the corresponding actions (for $N=1$) are 
$$(U^if)(\hat x^j)=e^{i\hat x^i}f(\hat x^j)$$ and 
$$(\hat U^if)(\hat x^j)=e^{i\hat x^i}f(\hat x^j+\theta^{ij}+f^{ij}_{\ \ k}\hat x^k).$$ 
Therefore a simple computation gives
$$(U^i\hat U^jf)(\hat x^l)=(U^ie^{i\hat x^j}f)(\hat x^l+\theta^{jl}+f^{jl}_{\ \ k}\hat x^k)=e^{i\hat x^i}e^{i\hat x^j}f(\hat x^l+\theta^{jl}+f^{jl}_{\ \ k}\hat x^k),$$
while 
$$(\hat U^j U^if)(\hat x^l)=(\hat U^je^{i\hat x^i}f)(\hat x^l)=e^{i\hat x^j}e^{i\hat x^i+i\theta^{ji}+if^{ji}_{\ \ k}\hat x^k}f(\hat x^l+\theta^{jl}+f^{jl}_{\ \ k}\hat x^k).$$
It is now easy to see that the Baker-Campbell-Hausdorff formula leads to the 
 desired commutation between $U^i$ and $\hat U^i$. Note that 
had we just computed the commutator between $\hat x^i$ and $\hat y^i$ we would have obtained an incorrect 
result, having missed the specific way that these operators act on states of the Hilbert space. This 
subtlety should be kept in mind for all similar computations involving $U$-operators in this paper.
 }. Moreover, they satisfy the commutation relations of a dual non-commutative twisted torus, namely 
\bea\label{tt3hatub}
\hat U^i\hat U^j&=&e^{ i\theta^{ij}+iNf_{\ k}^{ij}\hat x^k}\hat U^j\hat U^i.\eea
As before, substituting the solution back to the original matrix model action we obtain a gauge theory 
on this dual non-commutative twisted torus. We will return to this point later in this paper. 
\item Finally, the algebra of the phase space is determined to be 
\bea 
[\hat x^i, \hat x^j]&=& i\theta^{ij}+iNf^{ij}_{\ \ k}\hat x^k,\nn \\
{[}\hat p_i,\hat p_j]&=&0,\nn\\
{[}\hat p_i,\hat x^j]&=&-i\d_i^j-iNf_i^{\ jk}\hat p_k. \label{nctt3gen}
\eea 
It is directly observed that the momenta remain commutative, however the (previously constant) 
non-commutativity of the coordinates acquires a non-constant part. This is exactly a (nilpotent) Lie algebra 
type non-commutativity. 
\end{itemize}

Needless to mention that the formulae of the present paragraph are general, in the sense that they 
retain the same form for the compactification on any higher-dimensional nilmanifold. For example, a
 six-dimensional 
case was explicitly presented in  \cite{Chatzistavrakidis:2012yp}, where all the necessary data for a large 
class of higher-dimensional nilmanifolds may be found as well (see also Ref. \cite{Chatzistavrakidislie}).

A final comment on the case of twisted tori regards the Connes-Douglas-Schwarz correspondence. In the 
present case this translates into the statement that the non-constant non-commutativity of the coordinate 
operators, controlling also the non-commutative properties of the operators $U^i$ and $\hat U^i$, 
corresponds to turning on a geometric flux in  11-dimensional/type IIA supergravity 
\cite{Hulltts,prezasg2,Hullttm}. Schematically 
this means that 
\be 
\theta^{ij}(\hat x) \quad \overset{CDS}\longleftrightarrow \quad f_{ij}^{\ \ k}~
\ee 
This relation is supported by the resulting theory, as we will discuss in section 4.

\section{Non-geometric compactifications of Matrix theory}

In the previous section we remembered some of the established compactifications of Matrix theory on 
non-commutative tori and twisted tori. We already mentioned that these correspond to compactifications 
in the presence of a background B-field or geometric fluxes respectively. The natural question which 
immediately arises regards the implementation of other types of fluxes in this scheme. The most standard 
one is of course the NS-NS flux, which from the point of view of twisted tori  corresponds to a T-dual 
background with respect to the geometric flux one. However, it is by now well-known that there exist 
backgrounds with non-geometric fluxes formally obtained by further T-dualities of the geometric flux background. 
In this section we are going to describe Matrix theory flux compactifications with NS-NS $H$ flux, as well 
as non-geometric $Q$ and $R$ fluxes.  

Let us begin with the following observation. In the compactifications of section 2 it is evident that 
apart from non-commutative coordinate operators it is necessary to consider momentum operators as well. 
This leads naturally to a phase space description, reminiscent of the quantum-mechanical structure 
of Matrix theory. However, in section 2 we did not introduce the unitary operators 
$$\tilde U_i=e^{i\hat p_i},$$
which are the counterparts of the operators $U^i$ in the phase space. Of course, the reason that they were 
not introduced previously is that in the case of the twisted torus they commute among themselves,
 as well as with the $\mc X_i$ and 
therefore they did not play any crucial role up to now. In the present section we introduce these operators in 
order to exploit the full potential of the phase space. 

According to the above, there is now a second set of unitary operators acting on $\mc X_i$. This  
reminds us of the doubled formalism, which was used in order to construct the so-called  twisted doubled tori 
and provide an adequate description of non-geometric situations in terms of the geometry of an enlarged
 space. This realization motivates us to introduce also a set of ``dual'' Hermitian matrices $\tilde{\mc X}^i$ 
and impose on the matrices $\mc X_i$ and $\tilde{\mc X}^i$ the conditions which describe a compactification 
on a twisted doubled torus. Subsequently, we will utilize this formulation in order to describe 
the non-geometric compactifications and we will discuss the structure of the resulting theories.

\subsection{Algebraic building blocks for fluxes}

Here we are going to define and solve the conditions corresponding to a flux compactification of Matrix theory 
with $f, H, Q$ and $R$ flux. Technically the first one was already done in section 2, but  we shall 
reformulate it in terms of twisted doubled tori and in a way it will serve as a calibration of our formulation.

In all the following cases our solutions will  have the form
\bea 
\mc X_i &=& i\hat\partial_i + \hat{\mc A}_i, \nn\\  
\tilde{\mc X}^i &=& (-1)^{c_i}\hat x^i + \hat{\tilde{\mc A}}^i,
\label{genericx}\eea 
where the hat over the $\mc A$ and $\tilde{\mc A}$ denotes that these components may have to be 
appropriately corrected. In the second line we introduced a grading $(-1)^{c_i}$, which has to be 
included in order to guarantee that the Heisenberg relation is not spoiled. Its values will be given
 case by case in the following analysis. 
Moreover,  the unitary operators will always have the form 
\bea
U^i &=& e^{i\hat x^i}, \nn\\
\tilde U_i &=& e^{(-1)^{c_i}\hat\partial_i},\label{genericu}
\eea
with the same grading as above.
Finally, in the present section we set the constant part of the non-commutativity to zero, $\theta^{ij}=0$. 
These parameters may be reintroduced if necessary.

\paragraph{The $f$-block.}

This is the geometric flux situation, corresponding to a specific twisted doubled torus as described in
 \cite{hulltdt,prezastdt}. 
Here we follow the conventions of \cite{prezastdt}. The conditions that define the specific compactification 
 in the matrix model framework are the following: 
\bea 
U^i\mc X_i (U^i)^{-1} &=& \mc X_i + 1, \nn\\
U^1\mc X_3 (U^1)^{-1} &=& \mc X_3 - \mc X_2, \nn\\
U^2\mc X_3 (U^2)^{-1} &=& \mc X_3 + \mc X_1, \label{fc1}
\eea
and
\bea
\tilde U_i\tilde{\mc X}^i (\tilde U_i)^{-1} &=& \tilde{\mc X}^i + 1, \nn\\
\tilde U_3\tilde{\mc X}^1 (\tilde U_3)^{-1} &=& \tilde{\mc X}^1 + \mc X_2, \nn\\
\tilde U_3\tilde{\mc X}^2 (\tilde U_3)^{-1} &=& \tilde{\mc X}^2 - \mc X_1, \nn\\
U^1\tilde{\mc X}^2(U^1)^{-1} &=& \tilde{\mc X}^2+\tilde{\mc X}^3, \nn\\
U^2\tilde{\mc X}^1(U^2)^{-1} &=& \tilde{\mc X}^1-\tilde{\mc X}^3,
\eea
while all the combinations that do not appear in the above equations are supposed to have the 
trivial form $U\mc X U^{-1}=\mc X$ and similarly for the tilded ones. We directly observe that the first 
conditions, Eq. (\ref{fc1}), are the same that we encountered in section 2 for the twisted torus. 

When the unitary operators have the form (\ref{genericu}), then the connections appearing in Eq. 
(\ref{genericx}) solve all the compactification conditions under the following requirements:
\begin{itemize}
\item The grading exponent $c_i$ has the value 1 for $i=3$ and it is zero for $i\ne 3$.
 \item The algebra of $\{\hat x^i,\hat\partial_i\}$ has the form
\bea 
[\hat x^i, \hat x^j]&=& if^{ij}_{\ \ k}\hat x^k,\nn \\
{[}\hat \partial_i,\hat \partial_j]&=&0,\nn\\
{[}\hat \partial_i,\hat x^j]&=&\d_i^j-if_i^{\ jk}\hat \partial_k, \label{falg}
\eea
where only $f^{12}_{\ \ 3}=-f^{21}_{\ \ 3}=1$ are non-vanishing.
This is identical to the algebra we determined for the single twisted torus in Eq. (\ref{nctt3gen}) 
for $\theta^{ij}=0$ and $N=1$, 
as expected.
\item The gauge fields $\hat{\mc A}_i$ and $\hat{\tilde{\mc A}}^i$ are given by
\bea \label{gff}
\hat{\mc A}_i &=& {\mc A}_i+if_i^{\ jk}{\mc A}_j\hat \partial_k, \nn\\
\hat{\tilde{\mc A}}^i &=& \tilde{\mc A}^i-f_j^{\ ik}{{\mc A}_k}\hat x^j.
\eea
Note that $\mc A_i$ are functions of a set of operators $\hat U^i$ with the property $[\hat U^i,U^i]=0$.
\end{itemize}

The above requirements fix the $U$- and $\tilde U$-algebra and moreover allow us to determine the 
operators $\hat U^i$ and their algebra. Indeed we find 
\bea 
U^iU^j&=&e^{-if^{ij}_{\  k}\hat x^k}U^jU^i=f^{ij}_{\ k}(U^k)^{-1}U^jU^i, \nn\\
\tilde U_i\tilde U_j &=& \tilde U_j\tilde U_i, \nn\\
U^i\tilde U_j &=& e^{-(-1)^{c_j}f_j^{\ ik}\hat\partial_k}\tilde U_j U^i.\label{dfalg}
\eea
Moreover, the $\hat U^i$ and their algebra are identical to the ones appearing in Eqs. (\ref{gentwyy}) and 
(\ref{tt3hatub}) respectively, with $\theta^{ij}=0$. 

Let us now make an important observation concerning the phase space algebra (\ref{falg}).  It 
is simple to check that it is in fact non-associative. 
One way to see this is by calculating the double commutators 
\be 
-1=[[\hat p_3,\hat x^1],\hat x^2] \ne [\hat p_3,[\hat x^1,\hat x^2]]=1.
\ee
Alternatively, the Jacobiator with one $\hat p$ and two $\hat x$ entries does not vanish: 
\be \label{naf}
[\hat p_i, \hat x^j, \hat x^k] \equiv [[\hat p_i,\hat x^j],\hat x^k] + \mbox{c.p.} = -3f^{jk}_{\ \ i}.
\ee
In our framework this is a quantum-mechanical property, since it is a consequence of the Heisenberg relation 
between the coordinate and momentum operators. Upon introducing the quantum of action $\hbar$, it will 
appear multiplicatively on the right hand side of (\ref{naf}). Thus, as $\hbar\rightarrow 0$ this 
non-associativity will go away. Moreover, let us stress that the dynamical degrees of freedom, 
$\mc X$ and $\mc A$, do not exhibit this type of non-associativity. Further discussion on this point 
is left for section 3.4. 

As we already mentioned the extended structure introduced here does not add much to the results of 
the single twisted torus. Indeed, projecting to the relevant sector of $\mc X_i$ we obtain the 
geometric flux compactification of Matrix theory. However, the present formulation turns out to be very useful 
in order to account for other types of fluxes, as we will immediately do.

\paragraph{The $H$-block.}

In this paragraph we are going to describe a doubled compactification which will capture the case 
of a torus with $H$ flux. Such a compactification was described in \cite{hulltdt,prezastdt} in 
the supergravity picture. Along the lines of the $f$-block, this twisted doubled torus compactification 
is described in the matrix model with the following non-trivial  conditions on $\mc X_i$ and $\tilde{\mc X}^i$:
\bea 
U^i\mc X_i (U^i)^{-1} &=& \mc X_i + 1, \nn\\
\tilde U_i\tilde{\mc X}^i (\tilde U_i)^{-1} &=& \tilde{\mc X}^i + 1,
 \label{hc1}
\eea
and
\bea
U^i\tilde{\mc X}^j ( U^i)^{-1} &=& \tilde{\mc X}^j + H^{ijk}\mc X_k, 
\eea
where $H^{ijk}$ is fully antisymmetric and $H^{123}=1$.

When the unitary operators have the form (\ref{genericu}), we obtain a full solution as in Eq. 
(\ref{genericx})  under the following requirements:
\begin{itemize}
\item Concerning the grading exponent: $c_i=0$ for every $i=1,2,3$.
 \item The algebra of $\{\hat x^i,\hat\partial_i\}$ has the form
\bea 
[\hat x^i, \hat x^j]&=& H^{ijk}\hat \partial_k,\nn \\
{[}\hat \partial_i,\hat \partial_j]&=&0,\nn\\
{[}\hat \partial_i,\hat x^j]&=&\d_i^j. \label{halg}
\eea
\item The gauge fields $\hat{\mc A}_i$ and $\hat{\tilde{\mc A}}^i$ are given by
\bea 
\hat{\mc A}_i &=& {\mc A}_i, \nn\\
\hat{\tilde{\mc A}}^i &=& {\tilde{\mc A}}^i+iH^{ijk}{{\mc A}}_j\hat \partial_k,
\eea
i.e. $\mc A_i$ is not modified.
Note that, as always, the $\mc A_i$ are functions of a set of operators $\hat U^i$ with the property 
$[\hat U^i,U^i]=0$.
\end{itemize}

According to the above, the algebras of $U^i$ and $\tilde U^i$ are now fixed to be
\bea 
U^iU^j&=&e^{-H^{ijk}\hat \partial_k}U^jU^i=H^{ijk}(\tilde U^k)^{-1}U^jU^i, \nn\\
\tilde U_i\tilde U_j &=& \tilde U_j\tilde U_i, \nn\\
U^i\tilde U_j &=& \tilde U_j U^i.\label{gaH}
\eea
Moreover the operators $\hat U^i$, which centralize the $U^i$ ones, are
\bea
\hat U^i&=&e^{ix^i+iH^{ijk}\hat \partial_j\hat \partial_k},\eea
and they satisfy the dual relations  
\bea
\hat U^i\hat U^j&=&e^{H^{ijk}\hat \partial_k}\hat U^j\hat U^i=(H^{ijk}\tilde U^k)\hat U^j\hat U^i.
\eea
This non-commutativity should be related to a non-vanishing background B-field. 
Indeed, the commutativity of the gauge algebra is now obstructed by the parameters 
$$\theta^{ij}=H^{ijk}\hat p_k.$$
Then, the relation $\theta^{-1} \sim B$ dictates that the corresponding IIA supergravity is compactified 
on a torus with non-constant B field: 
\be 
B = x^1dx^2\wedge dx^3+x^2dx^3\wedge dx^1+x^3dx^1\wedge dx^2, 
\ee
where $x^i$ are the toroidal coordinates. This is in accord with the results of Ref. \cite{prezastdt}.

A similar observation to the previous case is that the phase space algebra (\ref{halg}) 
is also non-associative. 
Indeed, the Jacobiator with three $\hat x$ entries does not vanish: 
\be 
[\hat x^i, \hat x^j, \hat x^k]  = 3H^{ijk}.
\ee
It is very welcome that the Jacobiator is obstructed exactly by the object $H$, which is associated to the 
NS-NS flux. Note again that the dynamical variables $\mc X_i$ do not feel this non-associativity (see 
also section 3.4).

\paragraph{The $Q$-block.}

A further possibility arises when we consider the twisted doubled torus of Ref. \cite{prezastdt} which is 
related to a non-geometric $Q$ flux. This translates in the matrix model into the following 
non-trivial compactification conditions:
\bea 
U^i\mc X_i (U^i)^{-1} &=& \mc X_i + 1, \nn\\
U^1\mc X_2 (U^1)^{-1} &=& \mc X_2 + \tilde {\mc X}^3, \nn\\
U^1\mc X_3 (U^1)^{-1} &=& \mc X_3 -\tilde{\mc X}^2, 
 \label{qc1}
\eea
and
\bea
\tilde U_i\tilde{\mc X}^i (\tilde U_i)^{-1} &=& \tilde{\mc X}^i + 1,\nn\\
\tilde U_2{\mc X}_3 ( \tilde U_2)^{-1} &=& {\mc X}_3 + \mc X_1, \nn\\
\tilde U_3{\mc X}_2 ( \tilde U_3)^{-1} &=& {\mc X}_2 - \mc X_1, \nn\\
 \tilde U_2\tilde{\mc X}^1 ( \tilde U_2)^{-1} &=& \tilde{\mc X}^1 - \tilde {\mc X}^3, \nn\\
\tilde U_3\tilde {\mc X}^1 ( \tilde U_3)^{-1} &=& \tilde {\mc X}^1 + \tilde {\mc X}^2, 
\eea

With the unitary operators of Eq. (\ref{genericu}) and the solution of Eq. (\ref{genericx}),
 the corresponding requirements are:
\begin{itemize}
\item The grading exponent has the values $c_2=c_3=1$ and $c_1=0$.
 \item The algebra of $\{\hat x^i,\hat\partial_i\}$ has the form
\bea 
[\hat x^i,\hat x^j] &=& 0, \nn\\
{[}\hat \partial_i,\hat\partial_j] &=& Q_{ij}^{\ \ k}\hat\partial_k , \nn\\ 
{[}\hat\partial_i,\hat x^j]&=&\delta_i^j-Q_{ik}^{\ \ j}\hat x^k .
\label{Qalg}\eea
where $Q_{ij}^{\ \ k}$ is antisymmetric in its lower two indices and the only non-vanishing index 
structure is $Q_{23}^{\ \ 1}=1$. We directly observe the novel feature that the momenta are now non-commutative, while presently the
 coordinates 
commute among themselves.
\item The gauge fields $\hat{\mc A}_i$ and $\hat{\tilde{\mc A}}^i$ are given by
\bea 
\hat{\mc A}_i &=& {\mc A}_i+iQ_{ij}^{\ \ k}\tilde{\mc A}_j\hat \partial^k, \nn\\
\hat{\tilde{\mc A}}^i &=& \tilde{\mc A}^i+Q_{jk}^{\ \ i}\tilde{\mc A}^k\hat x^j.
\eea
\end{itemize}

The $U$- and $\tilde U$-algebra now becomes
\bea 
U^iU^j&=&U^jU^i, \nn\\
\tilde U_i\tilde U_j &=& e^{Q_{ij}^{\ \ k}\hat\partial_k}\tilde U_j\tilde U_i=(Q_{ij}^{\ \ k}\tilde U_k)
\tilde U_j\tilde U_i, \nn\\
U^i\tilde U_j &=& e^{-iQ_{jk}^{\ \ i}\hat x^k}\tilde U_j U^i,
\eea
where we observe that the operators $U^i$ now commute among themselves. This obviates the need to 
introduce the corresponding hatted operators $\hat U^i$. On the other hand, unlike the previous cases, 
now the $\tilde U_i$ operators do not commute among themselves. Therefore, for later use, it is 
instructive to introduce the hatted operators which commute with them, i.e. 
$[\hat{\tilde U}_i,\tilde U_j]=0$.
These operators read as
\bea
\hat{\tilde{ U_i}}=e^{(-1)^{c_i}\hat \partial_i-Q_{ij}^{\ \ k}\hat x^j\hat \partial_k}
\eea
and they satisfy an algebra dual to the one of $\tilde U_i$, namely
\be 
\hat{\tilde U}_i\hat{\tilde U}_j = e^{-Q_{ij}^{\ \ k}\hat\partial_k}\hat{\tilde U}_j\hat{\tilde U}_i=Q_{ij}^{\ \ k}(\hat{\tilde U}_k)^{-1}
\hat{\tilde U}_j\hat{\tilde U}_i.
\ee

As before, there is a non-associativity related to the algebra of the phase space. 
The relevant Jacobiator in the present case involves one $\hat x$ and two $\hat p$ entries and it reads as 
\be 
[\hat x^i, \hat p_j, \hat p_k]  = 3 Q^i_{\ jk}.
\ee
We observe that the Jacobiator is now obstructed exactly by $Q$. A novel feature is that here the 
dynamical matrices $\mc X_i$ do not commute among themselves. This will be discussed in the  subsection 3.3.

\paragraph{The $R$-block.}

The last algebraic block in this series should be related to the $R$ flux, which is captured by the 
appropriate twisted doubled torus. The compactification conditions implementing the geometry of 
this torus now read as
\bea 
U^i\mc X_i (U^i)^{-1} &=& \mc X_i + 1, \nn\\
\tilde U_i\tilde{\mc X}^i (\tilde U_i)^{-1} &=& \tilde{\mc X}^i + 1,
 \label{rc1}
\eea
and
\bea
\tilde U_i{\mc X}_j ( \tilde U_i)^{-1} &=& {\mc X}_j + R_{ijk}\tilde{\mc X}_k.
\eea

The usual requirements which fix the solution now are:
\begin{itemize}
\item Grading exponent: $c_i=1$ for all $i=1,2,3$.
 \item The algebra of $\{\hat x^i,\hat\partial_i\}$ has the form
\bea 
[\hat x^i,\hat x^j] &=& 0, \nn\\
{[}\hat \partial_i,\hat\partial_j] &=& -iR_{ijk}\hat x^k , \nn\\ 
{[}\hat\partial_i,\hat x^j]&=&\delta_i^j .
\label{Ralg}\eea
\item The gauge fields $\hat{\mc A}_i$ and $\tilde{\hat{\mc A}}^i$ are given by
\bea 
\hat{\mc A}_i &=& {\mc A}_i-R_{ijk}\tilde{\mc A}^j\hat x^k, \nn\\
\tilde{\hat{\mc A}}^i &=& \tilde{{\mc A}}^i .
\eea
\end{itemize}

The algebra of $U^i$ and $\tilde U_i$ operators is now found to be
\bea 
U^iU^j&=&U^jU^i, \nn\\
\tilde U_i\tilde U_j &=& e^{-iR_{ijk}\hat x^k}\tilde U_j\tilde U_i, \nn\\
U^i\tilde U_j &=& \tilde U_j U^i.
\eea
Similarly to the $Q$-block, the $U^i$ commute among themselves, unlike the $\tilde U^i$. Therefore 
we introduce the operators centralizing the latter, i.e.
\bea
\hat{\tilde{U_i}}=e^{-\hat \partial_i-iR_{ijk}\hat x^j\hat x^k},
\eea
which satisfy the dual algebra
\be 
\hat{\tilde U}_i\hat{\tilde U}_j = e^{iR_{ijk}\hat x^k}\hat{\tilde U}_j\hat{\tilde U}_i.
\ee 
Finally, the non-associativity related to the phase space algebra may be traced in the
 Jacobiator involving three $\hat p$ entries. Indeed we compute 
\be 
[\hat p_i, \hat p_j, \hat p_k]  = 3 R_{ijk},
\ee
thus obtaining an obstruction by $R$. Moreover, note that in the present case the
dynamical matrices $\mc X_i$ not only do not commute among themselves but they do not 
associate as well. This is part of the discussion which follows in subsection 3.3.

\subsection{Moving from block to block - T-duality}

The four algebras which we obtained for each flux-block appear in Eqs. (\ref{falg}),(\ref{halg}),(\ref{Qalg}) 
and (\ref{Ralg}). Let us now explore how starting with one of them we can obtain all the rest with 
appropriate transformations. Our starting point is the $H$-block which moreover is the leftmost entry in the 
T-duality chain (\ref{chain}). It is directly observed that the algebra (\ref{falg}) of the $f$-block 
is obtained by the $H$-block algebra (\ref{halg}) under the canonical transformation
\begin{align}
\hat x^3 &~ \rightarrow ~ -\hat p_3, \nn\\
\hat p_3 &~ \rightarrow ~\hat x^3.
\end{align}
Equivalently, we may rewrite this transformation as 
\bea \left(\begin{array}{ccc}
		\hat x^3  \\
		\hat p_3  \\
\end{array}\right) \rightarrow \left(\begin{array}{ccc}
		0 & -1  \\
		1 & 0  \\
\end{array}\right)\left(\begin{array}{ccc}
		\hat x^3  \\
		\hat p_3  \\
\end{array}\right)=-i\sigma_2\left(\begin{array}{ccc}
		\hat x^3  \\
		\hat p_3  \\
\end{array}\right).
\eea
Similarly, we can move from the algebra of the $f$-block to the algebra (\ref{Qalg}) of the $Q$-block 
by means of the canonical transformation
\begin{align}
\hat x^2 &~ \rightarrow ~ -\hat p_2, \nn\\
\hat p_2 &~ \rightarrow ~\hat x^2,
\end{align}
while the transformation from the $Q$-block algebra to the $R$-block one is similarly given by
\begin{align}
\hat x^1 &~ \rightarrow ~ -\hat p_1, \nn\\
\hat p_1 &~ \rightarrow ~\hat x^1.
\end{align}
Defining the six-dimensional column vector
$$
\hat q^i:=\left(\begin{array}{ccc}
		\hat x^i  \\
		\hat p_i  \\
\end{array}\right),
$$ 
the above transformations are realized by the following six-dimensional matrices 
\bea 
M_{H\rightarrow f}=\left(\begin{array}{cccccc}
1&0&0&0&0&0\\
 0&1&0&0&0&0 \\
0&0&0&0&0&-1\\
0&0&0&1&0&0\\
0&0&0&0&1&0\\
0&0&1&0&0&0
\end{array}\right), M_{f\rightarrow Q}=\left(\begin{array}{cccccc}
1&0&0&0&0&0\\
 0&0&0&-1&0&0 \\
0&0&1&0&0&0\\
0&0&0&1&0&0\\
0&1&0&0&0&0\\
0&0&0&0&0&1
\end{array}\right), \label{ct1} \eea 
and 
\bea
M_{Q\rightarrow R}=\left(\begin{array}{cccccc}
0&0&0&-1&0&0\\
 0&1&0&0&0&0 \\
0&0&1&0&0&0\\
1&0&0&0&0&0\\
0&0&0&0&1&0\\
0&0&0&0&0&1
\end{array}\right) \label{ct2}
\eea
respectively. Additionally, one may define the matrix which directly connects the starting $H$-block to 
the $R$-block. This is just the multiplication of the above three matrices and it is given as
\be 
M_{H\rightarrow R}=\left(\begin{array}{cccccc}
0&0&0&-1&0&0\\
 0&0&0&0&-1&0 \\
0&0&0&0&0&-1\\
1&0&0&0&0&0\\
0&1&0&0&0&0\\
0&0&1&0&0&0
\end{array}\right).\label{ct3}
\ee 
The above phase space transformations connect the algebraic blocks in the way we discussed. On the other 
hand we know that the solutions corresponding to each block should be related among themselves by T-dualities, in the spirit of the chain (\ref{chain}). However, it is evident that the above matrices 
are not elements of the T-duality group. Thus, these canonical transformations are not directly 
associated to T-dualities. At this point the grading operator that we introduced in (\ref{genericx}) 
comes into play. Let us consider the unitary operators $U^i$ and $\tilde U_i$ on equal footing by 
the column 
$$
\left(\begin{array}{ccc}
		U^i  \\
		\tilde U_i  \\
\end{array}\right)=\left(\begin{array}{ccc}
		e^{i\hat x^i}  \\
		e^{(-1)^{c_i}i\hat p_i}  \\
\end{array}\right):=e^{(-1)^{\hat c_i}i\hat q^i},
$$ 
where we introduced the grading operator $(-1)^{\hat c_i}$, representing the action of $(-1)^{c_i}$ on all the 
unitary operators.
If we represent this grading operator by a six-dimensional matrix 
acting on the column vector 
$\hat q^i$,
then for the $H$-block it is just 
$$(-1)^{\hat c_i}_{H}=\one_6,$$
where $\one_6$ denotes the unit matrix in six dimensions. For the $f$-block this operator has the 
diagonal form
\be 
(-1)^{\hat c_i}_{f}=\mbox{diag}(1,1,1,1,1,-1),
\ee 
and similarly for the other two blocks,
\bea 
(-1)^{\hat c_i}_{Q}&=&\mbox{diag}(1,1,1,1,-1,-1), \\
(-1)^{\hat c_i}_{R}&=&\mbox{diag}(1,1,1,-1,-1,-1).
\eea 
Therefore we now have at hand two operations, namely the canonical transformations given by the matrices $M$, 
which connect just the algebraic building blocks, 
and the gradings $(-1)^{\hat c_i}$, which preserve the Heisenberg relation. When we move from one block to the 
other it is the combined action of these two operations that connects one solution to another. 
 Let us consider for 
example the case of moving from the $H$- to the $f$-block. The corresponding operation is 
\be 
M_{H\rightarrow f}\cdot(-1)^{\hat c_i}_f=\left(\begin{array}{cccccc}
1&0&0&0&0&0\\
 0&1&0&0&0&0 \\
0&0&0&0&0&1\\
0&0&0&1&0&0\\
0&0&0&0&1&0\\
0&0&1&0&0&0
\end{array}\right):=T_3.
\ee 
The latter, $T_3$, is an element of the compact subgroup of the T-duality group and generates (at least formally) a 
T-duality along the $x^3$ direction of the torus, thus leading to the corresponding twisted torus 
without $H$ flux. This way one can explain the T-duality chain (\ref{chain}) in the present framework.

The above results can be represented as in the following diagram:
\[\renewcommand{\arraystretch}{2.0}
\begin{array}{ccccccc}
H & \overset{T_3}\longleftrightarrow & f& \overset{T_2}\longleftrightarrow &
 Q & \overset{T_1}\longleftrightarrow & R\\
\Biggl\updownarrow& & \Biggl\updownarrow  && \Biggl\updownarrow && \Biggl\updownarrow
\\
\theta(\hat p)&\overset{M_{H\rightarrow f}\cdot (-1)^{\hat c_i}_f}\longleftrightarrow&\theta(\hat x)& \overset{M_{f\rightarrow Q}\cdot (-1)^{\hat c_i}_Q}\longleftrightarrow& \tilde\theta(\hat p) & \overset{M_{Q\rightarrow R}\cdot (-1)^{\hat c_i}_R}\longleftrightarrow & \tilde\theta(\hat x)
\end{array}\]
The vertical arrows 
denote the CDS correspondence, i.e. they connect a compactification of the matrix model to a compactification 
of low-energy supergravity. The upper horizontal arrows denote T-duality among supergravity vacua, 
while the lower horizontal arrows denote the operations in the matrix model which correspond to these T-dualities.
 The matrix model solutions are represented by the non-commutativity parameters
\bea 
\theta^{ij}&=&[\hat x^i, \hat x^j], \\
\tilde\theta_{ij}&=&[\hat p_i, \hat p_j],
\eea 
whose dependence appears in the parentheses. 

Finally, it is worth mentioning the case when one moves directly from the $H$-block to the $R$-block. According 
to the above rule, the T-duality should be achieved by the operation 
\be 
M_{H\rightarrow R}\cdot(-1)^{\hat c_i}_R=\left(\begin{array}{cccccc}
0&\one_3\\
 \one_3&0 \\
\end{array}\right),
\ee 
which is indeed also an element of the T-duality group.

\subsection{Position/Momentum space duality}

The doubled formulation of section 3.1 motivated the introduction of the additional Hermitian matrices 
$\tilde {\mc X}^i$, which facilitated the implementation of twisted doubled tori in the present framework. 
However, these matrices should not be really dynamical degrees of freedom, since they are not present 
in Matrix theory\footnote{However, note that they could be thought of as coexisting sectors of Matrix 
theory, i.e. different multi-brane solutions which are combined block-diagonally as e.g. in 
Ref. \cite{Chatzistavrakidisinter}. Moreover, it is conceivable that $\tilde {\mc X}^i$  may be part of a ``doubled 
matrix model'', which could serve as a non-perturbative definition of double field theory \cite{dft}. 
We shall not explore further these possibilities in the present paper.}. Thus we would like here to restrict the 
previous formulation to the subsectors which contain only the true dynamical components. This procedure is 
similar to the so-called ``polarizations'' of \cite{hulltdt}.

Let us warm up this discussion with the case of the geometric flux $f$,
 which was presented both as a twisted torus in section 2.3 and as a twisted doubled torus in section 3.1.
As we already mentioned, the overlapping non-trivial compactification conditions and the phase space algebras in these 
two cases fully coincide, as well as the form of the dynamical matrices $\mc X_i$. This directly shows that there 
is a trivial projection from one to the other. In a way, the introduction of $\tilde{\mc X}^i$ is totally 
redundant in this case.

Turning to the $H$-block, which is related to an NS-NS flux background in the supergravity picture, 
it is again plausible to determine a well-defined projection to a single torus. Indeed, we can consider only 
the non-trivial conditions 
\be \label{condhsingle}
U^i\mc X_i(U^i)^{-1}=\mc X_i+1,
\ee 
dressed with the commutation relation 
\be \label{uh}
U^iU^j=e^{-H^{ijk}\hat\partial_k}U^jU^i
\ee
for the operators $U^i$. The conditions (\ref{condhsingle}) define a compactification of matrix theory on a 
torus, while the relation (\ref{uh}) indicates that the algebra of functions on this torus is deformed, 
its non-commutativity being controlled by $H^{ijk}$. We already discussed that this non-commutativity is 
related to the presence of a non-vanishing and non-constant B-field.

The next and more interesting situation is the $Q$-block. Can we define a projection from the twisted 
doubled torus to a single toroidal-like compactification? Or, alternatively, can we define a 
legitimate Matrix theory compactification based on the algebra (\ref{Qalg})? 
Suppressing for a moment the $\mc A_i$ part in the connections, let us examine what happens for the 
standard form of the solutions, i.e. when $\mc X_i=i\hat\partial_i$ and $U^i=e^{i\hat x^i}$. 
It is straightforward to obtain 
\be 
U^1\mc X_2 (U^1)^{-1} = \mc X_2 - \hat x^3.
\ee  
This is a bizarre relation which at first sight does not seem to have a clear interpretation apart 
from the one in the extended context of twisted doubled tori, where the $\hat x^3$ is actually 
$\tilde{\mc X}^3$. Furthermore, presently the operators $U^i$ commute among themselves, which makes 
the situation even more obscure.

In order to clarify the above situation let us make the following observations. Define the deformation 
parameters
\bea
\theta^{ij}|_f&=&f^{ij}_{\ \ k}\hat x^k, \\
\tilde \theta_{ij}|_Q&=&Q_{ij}^{\ \ k}\hat p_k,
\eea
in obvious notation. There is already a flavour of duality in these relations, which is intensified 
by looking at the $U/\tilde U$ relations:
\begin{align} 
f: \quad   U^iU^j(U^i)^{-1}(U^j)^{-1}&=f^{ij}_{\ \ k}(U^k)^{-1},   &\tilde U_i\tilde U_j=\tilde U_j\tilde U_i,  \\
Q: \quad  \tilde U_i\tilde U_j(\tilde U_i)^{-1}(\tilde U_j)^{-1}&=Q_{ij}^{\ \ k}\tilde U_k, & U^iU^j=U^jU^i,
\end{align}
It is then clear that the structure of $U^i$ in the $f$-block mimics the structure of $\tilde U_i$ in the 
$Q$-block and vice versa. Thus we can argue that the projection which defines a compactification of 
Matrix theory in the $Q$-block case is the one to the tilded subsector. In particular, we consider 
the tilded quantities $\tilde{\mc X}^i$ and $\tilde U_i$ with the conditions 
\bea 
\tilde U_i\tilde{\mc X}^i (\tilde U_i)^{-1} &=& \tilde{\mc X}^i + 1,\nn\\
 \tilde U_2\tilde{\mc X}^1 ( \tilde U_2)^{-1} &=& \tilde{\mc X}^1 - \tilde {\mc X}^3, \nn\\
\tilde U_3\tilde {\mc X}^1 ( \tilde U_3)^{-1} &=& \tilde {\mc X}^1 + \tilde {\mc X}^2.
\eea 
The dynamical matrices are now the $\tilde{\mc X}^i$ instead. This compactification is now the 
twisted torus compactification on the dual space to the one operating in the case of the geometric flux 
$f$. More specifically, let us recall that in the position representation of quantum mechanics the 
classical variables are mapped to Hermitian operators as 
\be
x \rightarrow \hat x, \quad p\rightarrow \hat p = -i\hbar\frac{\partial}{\partial x}, 
\ee
while in the momentum representation the correspondence is 
\be
x \rightarrow \hat x=i\hbar\frac{\partial}{\partial p}, \quad p\rightarrow \hat p. 
\ee
This simple observation shows that there is an exact correspondence between geometric $f$ flux in the 
position space and non-geometric $Q$ flux in the momentum space: 
$$
\theta^{ij}|_f \quad \mbox{in} \quad \hat x\mbox{-space} \quad \longleftrightarrow 
\quad \tilde\theta_{ij}|_Q \quad \mbox{in} \quad \hat p
\mbox{-space} ~.
$$

There is a final case which one would like to project, that of the $R$-block. As expected, this exhibits 
similar features to the $Q$-block case and it has to be interpreted as above. In particular, 
assuming the algebra (\ref{Ralg}) with
$\mc X_i=i\hat\partial_i$ and $U^i=e^{i\hat x^i}$,
we obtain 
\bea 
U^i\mc X_i (U^i)^{-1} &=& \mc X_i +1, \\
U^iU^j&=&U^jU^i.
\eea  
This might seem a totally well-defined compactification, i.e. a compactification on a standard torus 
such as the ones we referred to in section 2.2 with $\theta^{ij}=0$. However, the situation is not as 
simple as this since it turns out that in the present case the dynamical matrices $\mc X_i$ are 
non-associative:
\be 
[\mc X_i, \mc X_j,\mc X_k]\ne 0.
\ee
This is rather expected in view of the known properties of non-geometric compactifications
 \cite{ngncna1,ngncna2,ngncna3,Grange}, 
but it is awkward to think that a set of Hermitian matrices does not satisfy the Jacobi identity. 
In fact it is then impossible to represent them on a Hilbert space.
The discussion of the $Q$-block comes to the rescue, since we can instead consider the projection to the 
tilded sector, which satisfies the conditions
 \bea 
\tilde U_i\tilde{\mc X}^i (\tilde U_i)^{-1} = \tilde{\mc X}^i + 1, 
\eea
with
\bea 
\tilde U_i\tilde U_j = e^{R_{ijk}\hat x^k}\tilde U_j\tilde U_i. 
\eea
This comes on a dual footing to the $H$-block case. In particular, the relevant deformation parameters are
\bea
\theta^{ij}|_H&=&H^{ijk}\hat p_k, \\
\tilde \theta_{ij}|_R&=&R_{ijk}\hat x^k.
\eea
Thus there is an exact correspondence between NS-NS $H$ flux in the 
position space and non-geometric $R$ flux in the momentum space: 
$$
\theta^{ij}|_H \quad \mbox{in} \quad \hat x\mbox{-space} \quad \longleftrightarrow 
\quad \tilde\theta_{ij}|_R \quad \mbox{in} \quad \hat p
\mbox{-space} ~.
$$
In the momentum space the compactification related to the $R$-block is well-defined and moreover 
the dynamical matrices are associative.

Summarizing,
\begin{itemize}
 \item In position space, there are well-defined compactifications of Matrix theory with non-constant 
 non-commutativity $\theta^{ij}$ among the coordinate operators, which is related to the presence of geometric or NS-NS 
fluxes in the corresponding supergravity compactifications. The compactifications related to the $Q$- and 
$R$-algebras which exhibit non-commutativity among the momentum operators are not well-defined.
\item In momentum space, the compactifications with non-commutativity $\tilde \theta_{ij}$ among the momentum 
operators 
are well-defined and they correspond to supergravity with $Q$ and $R$ fluxes. The cases based on the 
$H$- and $f$-algebras are presently not well-defined. 
\end{itemize}
The situation we just described has a very similar incarnation in low-energy supergravity. The authors 
of Refs. \cite{Andriot1,Andriot2} studied non-geometric compactifications from the perspective of 
generalized geometry \cite{gg1,gg2}. They showed that while non-geometric configurations are ill-defined in a frame where the generalized 
metric is parametrized by the 
B-field $B_{ij}$, they become well-defined in another 
frame where the generalized metric is parametrized by the antisymmetric bivector $\beta^{ij}$ of 
generalized geometry. 
In the second frame, geometric configurations are instead ill-defined. This means that one has 
to choose the appropriate generalized vielbein which would yield the correct, i.e. well-defined, Lagrangian 
for each configuration.

The above discussion allows us to make the reasonable speculation that just as the 
deformation parameter $\theta$ is related to 
the B-field via
\be \label{thetab}
\theta^{ij} \sim (B_{ij})^{-1},
\ee  
there should exist a similar dependence among the deformation parameter $\tilde\theta$ and the bivector 
$\beta$ of generalized geometry as
\be \label{thetabeta}
\tilde \theta_{ij} \sim (\beta^{ij})^{-1}.
\ee 
The latter relation should be studied in detail by comparison of the corresponding effective actions. 

Moreover, let us remind that in supergravity $f$ is a (spin) connection and $H$ is a tensor field. 
The above duality between position and momentum space in the matrix model indicates that in 
momentum space $Q$ should instead be a connection and $R$ should be a tensor field. A similar 
observation was made in \cite{Andriot2}.

A final 
remark regards the index structure of the encountered quantities. Observe that the indices in the T-duality chain
(\ref{chain}) are in exactly the opposite position to the corresponding parameters in the phase space 
algebras. Recall that the former are related to the supergravity picture while the latter to the matrix model 
picture. Then the relations (\ref{thetab}) and (\ref{thetabeta}) explain the above index structure.

\subsection{Resolution of non-associativity and flux quantization}

Let us now make some important remarks regarding the non-assocativity we encountered above. 
First of all, we could say that there are two types of non-associativity which have to be 
treated differently. The first type is the one among the dynamical degrees of freedom $\mc X_i$. This 
issue was already discussed in the previous subsection within position-momentum space 
duality. We saw that in the appropriate representation, where the compactification is well-defined, there 
is no sign of non-associativity. In particular, for the $R$-type solution in momentum space the 
dynamical degrees of freedom are perfectly associative. 

However, there is a second type of non-associativity which we have not discussed in detail so far. 
This has a common origin to the first one, i.e. the phase space algebra, but it regards the 
non-dynamical algebraic elements $U^i$. Indeed, let us look at the $H$-block, with algebra 
(\ref{halg}) and commutation relation among the $U^i$ as in Eq. (\ref{uh}). It is strightforward 
to compute that
\be \label{una}
U^i(U^jU^k)=e^{\frac i2 H^{ijk}}(U^iU^j)U^k.
\ee 
This relation should be thought of as an anomalous 3-cocycle. Indeed, exactly the same relation may be 
found in \cite{Jackiw2}, where its appearance in a physical system is discussed. However, such a 
3-cocycle cannot be tolerated. In the words of Jackiw, {\it{``It is important to appreciate that 
non-associating quantities cannot be represented by well-defined linear operators, acting on a vector 
or Hilbert space, since by definition operators on vectors necessarily associate''}} (see page 19 of 
Ref. \cite{Jackiw2}). The resolution of this problem leads naturally to a Dirac quantization condition. 
Requiring associativity to be restored in Eq. (\ref{una}) we directly obtain 
\be 
H^{ijk}=4\pi n, \quad n\in\Z.
\ee 
Therefore we observe that the analog of the $H$ flux in the framework of Matrix theory has to obey 
a quantization condition. This is very plausible because in string theory fluxes have to be 
quantized \cite{Grana}.

Let us move on to the geometric flux and examine what happens in this case. Taking into account the 
algebra (\ref{falg}) and the commutation relation among the operators $U^i$, we obtain
\be 
U^i(U^jU^k)=(U^iU^j)U^k, 
\ee 
namely the present situation is already associative. This is expected, since the $f$ flux does not arise 
from a $p$-form source, but it is a metric flux.

As far as the $Q$ and $R$ cases are concerned, there is not much to add. As we already discussed above, they 
provide well-defined solutions in the momentum space, where they play the corresponding role of $f$ and $H$. 
Therefore, $R$ obeys a quantization condition much like $H$, while for $Q$ the corresponding 
operators already associate.

It should be noted that apart from the above operators, associativity should be guaranteed for the 
gauge fields $\mc A$ of each solution. These are functions of the hatted operators $\hat U^i$, 
which associate whenever the unhatted ones do. Therefore, the above discussion applies equally well for the 
gauge fields too.  

Let us mention that in the recent Ref. \cite{hz} and in the context of double field theory it is found that 
large gauge transformations do associate even in cases when the coordinate maps do not. 
It is notable that this exhibits a similarity 
to our present discussion in a rather different framework.

\subsection{Putting the blocks together}

It is well-known that in supergravity it is possible to consider compactifications where different types 
of fluxes coexist\footnote{Usually this is indispensable in order to obtain a true 
string vacuum solving the string equations of motion, which is not the case for the toy model of a 
twisted 3-torus. R-R fluxes are then important too, but we will not discuss them here.}.
 For example one can consider simultaneously geometric and NS-NS fluxes, i.e. a 
twisted torus penetrated by an $H$ flux. Therefore it is reasonable to ask whether the different algebraic 
blocks described above may be combined to yield Matrix theory compactifications with a 
superimposition of deformation parameters $\theta$ and/or $\tilde\theta$. This question boils down to 
the attempt of defining and solving the appropriate compactification conditions which would yield the 
general phase space algebra
\bea
[\hat x^i,\hat x^j] &=&if^{ij}_{ \ \ k}\hat x^k + H^{ijk}\hat\partial_k , \nn\\
{[}\hat \partial_i,\hat\partial_j] &=& Q_{ij}^{\ \ k}\hat\partial_k-iR_{ijk}\hat x^k , \nn\\ 
{[}\hat\partial_i,\hat x^j]&=&\delta_i^j-Q_{ik}^{\ \ j}\hat x^k- if^{jk}_{\ \ i}\hat\partial_k .
\label{fullalg}
\eea 
In the doubled formalism of section 3.1, the form of $\mc X_i, U^i$ and the corresponding tilded
 ones
leads to some compactification conditions which combine all the blocks that we discussed in section 3.1.
Although these conditions define some non-commutative twisted doubled torus, for Matrix theory we would like 
to have a projection on a single set of dynamical variables (see, however, footnote 3). Let us now discuss 
some of them.

A first possibility is to consider $\mc X_i$ as the Matrix theory variables. This choice includes the 
first two projections of section 3.3 but it also allows for the combination of the two. This 
amounts to the deformation parameters
\be \label{fH}
\theta^{ij}=H^{ijk}\hat p_k + f^{ij}_{\ \ k}\hat x^k, \quad \tilde\theta_{ij}=0.
\ee   
The first term in $\theta$ corresponds to a non-constant B-field, while the second term to a geometric 
flux. Thus, this situation would be related to a twisted torus with NS-NS flux in supergravity.

Equally well we could project to the $\tilde X^i$ sector, which would now provide the dynamical variables 
of Matrix theory. This choice includes the latter two projections of section 3.3, as well as their 
combination. As discussed above, both these projections are ill-defined in $\hat x$-space but they 
become well-defined in $\hat p$-space. The deformation parameters are now 
\be \label{QR}
\theta^{ij}=0,  \quad \tilde\theta_{ij}=Q_{ij}^{\ \ k}\hat p_k + R_{ijk}\hat x^k,
\ee 
and in $\hat p$-space they define a Matrix theory compactification analogous to a supergravity background 
with $Q$ and $R$ fluxes.

It is not clear how one could combine all the fluxes in a single case. One possibility would be to 
consider the sum of two twisted tori, $\tilde{\text{T}}^3_A\oplus \tilde{\text{T}}^3_B$ and associate the Hermitian 
matrices $\mc X_{i}$ to the first twisted torus and the matrices $\mc X_{i+3}$ to the second one, for 
$i=1,2,3$. Then one could think of using the solution related to the  Eq. (\ref{fH}) on the first torus 
and the solution associated to Eq. (\ref{QR}) on the second. This is a legitimate possibility but 
its interpretation is rather obscure and therefore we would not like to pursue it further in the present 
paper. 

\section{Remarks on the gauge theory}

In the previous section we determined several types of solutions of Matrix theory compactified on deformed 
tori. In the prototype example of a torus with constant non-commutativity, when the solution is substituted
back into the original action functional the effective action describes a 
non-commutative gauge theory coupled to scalars and fermions \cite{cds}. This resulting action was 
subsequently  compared to the action of a string moving in the background of a constant B-field. 
This comparison further supports the correspondence between non-commutativity parameters and 
background field values \cite{DouglasHull,Kawano,Chu}. 

Generically, the compactification of the matrix model on a standard 3-torus will lead to the following tree-level
effective action:
\bea\label{cT3}
{\cal S}_{\rm{eff}}\propto \int dt\;{\rm Tr}\left\{-\dot{\mathcal D}_i^2
+\dot{\mathcal A}_m^2+\frac 12 [{\mathcal D}_i,{\mathcal D}_j]^2-
i[{\mathcal D}_i,{\mathcal A}_m]^2-\frac 12[{\mathcal A}_m,{\mathcal A}_n]^2+{\rm fermions} 
\right\}. \nn
\eea
Here  we replaced the matrices ${\mathcal X}_i$ with their solution, namely with covariant derivatives.
Ignoring the scalar part of the action coming from $\mc X_m$ and defining $F_{ij}=[{\mathcal D}_i,{\mathcal D}_j]$ one obtains 
\be\label{gauge}{\cal S}\propto \int dt\;{\rm Tr}(F_{ij}F^{ij}).\ee

In the case of compactification on a non-commutative torus \cite{cds}, the action (\ref{gauge}) is defined on 
the dual non-commutative torus given by the relations (\ref{cteht}). It is possible to represent the action (\ref{gauge}) on the space of commuting variables and  
rewrite the trace over infinite-dimensional matrices as 
\be\label{Trint}
{\rm Tr}\to \int d^3 x\; {\rm tr},\ee
where $x^i$ are periodic coordinates on $\text {T}^3$ and ${\rm tr}$ denotes the trace over n-dimensional Hermitean matrices ${\mc A}_i$. Representing the matrices ${\mc A}_i$ as functions of commuting variables $A_i(x^j)$ one obtains  the 
non-commutative field strength in the form  
\be\label{fs} F_{ij}=\partial_i { A}_j-\partial_j { A}_i + i{ A}_i\star { A}_j-i{ A}_j\star { A}_i, \ee
where the Moyal-Weyl $\star$ product
\be 
f\star g = e^{\frac i2 \frac{\partial}{\partial x^i}\hat\theta^{ij}\frac{\partial}{\partial y^j}}f(x)g(y)|_{y\rightarrow x},
\ee 
encodes the non-commutativity of the algebra of functions (\ref{cteht}) on the non-commutative torus.

\subsection{Gauge theory with fluxes}

Let us now discuss some features of the gauge theory resulting from the compactifications with fluxes. In the following   we concentrate on the pure gauge sector of the compactified action.

In the case of geometric flux,  the solutions of the compactification conditions are:
\bea \label{stff}
{\mathcal X}_i=i{\mathcal D}_i &=& i\hat\partial_i+ {\mc A}_i(\hat U)+if_i^{\ jk}{\mc A}_j(\hat U)\hat \partial_k,\nonumber \\
{\tilde{\mathcal X}}^i=i{\tilde{\mathcal D}}^i &=& (-1)^{c_i}x^i+\tilde{\mc A}^i(\tilde U)-f_j^{\ ik}{{\mc A}_k(\hat U)}x^j.
\eea
Note  that ${\tilde {\mc A}}^i$ are polynomial functions of $\tilde U_k=\exp((-)^{c_k}\partial_k)$, while ${\mc A}_i$ depend on $\hat U^j=\exp(i\hat y^j)$, with $\hat y^j = \hat x^j-if^{ji}_{\ \ k}\hat x^k\hat\partial_i$.
The gauge transformations corresponding to the compactification of the original action on 
the twisted doubled torus are generated by $U^i$:
\bea\label{gtf}
U^i {\mc {\hat A}}_j (U^i)^{-1} &=& {\mc {\hat A}}_j-f_j^{\ ik}\mc {\hat A}_k,\nn \\
U^i \tilde{{\mc A}^j} (U^i)^{-1} &=& \tilde{{\mc A}^j}+f^{\ ij}_{k}\tilde{{\mc A}^k},
\eea
and $\tilde U_i$
\bea\label{gtfd}
\tilde U_i {\mc {\hat A}}_j (\tilde U_i)^{-1} &=& {\mc {\hat A}}_j,\nn \\
\tilde U_i \tilde{{\mc A}^j} (\tilde U_i)^{-1} &=& \tilde{{\mc A}^j}.
\eea
Let us recall that the corrected 
gauge field has the form $\mc {\hat A}_i = \mc A_i+if_i^{\ jk}\mc A_j\hat\partial_k$.
The algebra of these gauge generators, given in Eq. (\ref{dfalg}), represents in the matrix model 
framework the gauge algebra of 
the corresponding string compactification, which was determined in Ref. \cite{km}.
 
Inserting the  solutions  (\ref{stff}) back into the original action (\ref{BFSSaction}) leads to the following effective action:
\bea \label{effdual}
{\cal S}_{\rm{eff}}\propto \int dt\;{\rm Tr}\left(  [{\mathcal D}_i,{\mathcal D}_j]^2+ 2 [{\mathcal D}_i,\tilde{\mathcal D}^j]^2+ [\tilde{\mathcal D}^i,\tilde{\mathcal D}^j]^2\right).\eea
As we noticed before, the dual formulation is redundant in the case of geometric flux. The gauge (sub)algebra generated by $U^i$ closes, so   projecting on the physical torus we get 
\bea \label{effduall}
{\cal S}\propto \int dt\;{\rm Tr}   [{\mathcal D}_i,{\mathcal D}_j]^2=\int dt\;{\rm Tr} (F_{ij}F^{ij}),\eea
where the non-commutative field strength in the present case is defined in terms of $\hat {\mc A}_i$ as 
$F_{ij}=\partial_i \hat{\mc A}_j-\partial_j \hat{\mc A}_i - i[\hat{\mc A}_i,\hat{\mc A}_j]$.

Notice that the operators $\hat U^j$ generate the gauge transformations in the resulting gauge theory (\ref{effduall}). Moreover, one can show that these operators define the compactification on the dual torus:
\bea 
\hat U^i\mc X_i (\hat U^i)^{-1} &=& \mc X_i + 1, \nn\\
\hat U^1\mc X_3 (\hat U^1)^{-1} &=& \mc X_3 - 2\mc X_2, \nn\\
\hat U^2\mc X_3 (\hat U^2)^{-1} &=& \mc X_3 + 2\mc X_1, \label{fcd1}
\eea
with
\be\hat U^i {\mc {\hat A}}_j (\hat U^i)^{-1} = {\mc {\hat A}}_j.\ee 
This supports our claim that the non-constant non-commutativity of coordinate operators corresponds to turning on a geometric flux in supergravity.

Representing the action (\ref{effduall}) on the space of commuting coordinates is a non-trivial task.
 However, it is important to notice that the non-commutative structure of the algebra of  functions on the torus would be encoded using the  $\star$ product:
\be 
f\star g = e^{-\frac i2 f^{ij}_{\ \ k}x^k\frac{\partial}{\partial y^i}\frac{\partial}
{\partial z^j}}f(y)g(z)|_{y,z\rightarrow x}~,
\ee 
which is associative, as is the whole algebra of functions on the twisted torus.

Let us turn to the compactification incorporating the $H$ flux. This case resulted in the following solutions:
\bea \label{sth}
{\mathcal X}_i=i{\mathcal D}_i &=& i\hat\partial_i+ {\mc A}_i(\hat U),\nonumber \\
{\tilde{\mathcal X}}^i=i{\tilde{\mathcal D}}^i &=& x^i
+{\tilde{\mc A}}^i(\tilde U)+iH^{ijk}{{\mc A}_j(\hat U)}\partial_k.
\eea
The gauge transformations are generated by $U^i$ and $\tilde U_i$:
\bea\label{gth}
&& U^i {\mc A}_j (U^i)^{-1} = {\mc A}_j,\quad U^i \hat{\tilde{\mc A}}^j (U^i)^{-1} = \hat{\tilde{\mc A}}^j+H^{ijk}{\mc A}_k,\nn\\
&& \tilde U_i {\mc A}_j (\tilde U_i)^{-1}= {\mc A}_j,\quad\tilde U_i \hat{\tilde{\mc A}}^j (\tilde U_i)^{-1} = \hat{\tilde{\mc A}}^j.
\eea
The gauge algebra, given in (\ref{gaH}), is now more involved. Unlike the case of geometric flux, here the algebra of generators $U^i$ does not close. 
Inserting the solutions (\ref{sth}) in the original BFSS action we again obtain (\ref{effdual}), but in this case the second term under the integral will have a contribution to the projected action. We find that the projected action on the (plain) torus is of the form:
\bea\label{hact}
{\cal S}\propto \int dt\;{\rm Tr}\left(\frac{1}{4}F_{ij}^2+\frac{1}{2}H^{ijk}{\cal X}_k F_{ij}-H^{ijk}{\cal A}_k F_{ij}+\frac{i}{2}H^{ijk}[{\mc A}_i,{\mc A}_j]{\cal X}_k+{\cal O}(H^2)\right).\eea
The first H-dependent term is basically a Myers term \cite{Myers},  here obtained from the fluctuations without a 
need to modify the original action. 
Let us note that such a term was also obtained in Refs. \cite{cs1,cs2} from the 
expansion of the Dirac-Born-Infeld action. The second and the third H-dependent terms appear as (parts of) 
the Chern-Simons action. 
Let us recall that here we substituted the solutions in the tree-level matrix model action. 
It would be very interesting to determine the 1-loop effective action of the solutions we discussed 
and analyze the corresponding new terms. This task may be pursued along the lines of techniques 
used in Refs. \cite{bs,ts}.

Moreover, the gauge tranformations induced by $\hat U^i$ define a compactification  on the dual torus:
\be\hat U^i {\mc X}_j (\hat U^i)^{-1}= {\mc X}_j +1, \quad \hat U^i {\mc {\hat A}}_j (\hat U^i)^{-1} = {\mc {\hat A}}_j ,\ee 
showing that this matrix model compactification 
 is related to a supergravity compactification on a torus with NS-NS flux. 

The compactifications incorporating $Q$ and $R$ fluxes do not reveal any new structure. Defining the physical space as the one where the algebra of functions is associative, lead us to  the projection  on the dual momentum space. The effective tree-level gauge action in these cases will be dual to the ones for $f$ and $H$ cases
 as discussed in section 3.3.

\section{Conclusions}

The BFSS model or Matrix theory is a matrix model which, in its $N\rightarrow \infty$ limit, is conjectured 
to be equivalent to uncompactified M theory \cite{Banks:1996vh}. Thus, if Matrix theory indeed serves as a 
non-perturbative definition of M theory,  it should contain its low-energy limit, namely 
11-dimensional supergravity on flat spacetime. Moreover, compactifications of the matrix model should 
provide another description of compactified supergravity backgrounds. Progress towards this direction 
revealed that matrix compactifications on non-commutative tori are related to supergravity compactifications 
with background fields \cite{cds,DouglasHull}.

In the present paper we explored connections between compactifications of Matrix theory and flux compactifications 
of supergravity, with geometric, NS-NS and non-geometric fluxes. The quantum-mechanical nature of the BFSS model 
assigns an important role to the phase space. In particular, different non-commutative deformations of the
phase space algebra lead to certain solutions of the model and the parameters of the non-commutativity can be 
related to  fluxes in the corresponding supergravity compactification. Moreover, the algebraic building blocks 
for these solutions can be related via certain operations which provide a matrix model realization of the T-duality 
chain  
\be 
H_{ijk} \overset{T_k}\longrightarrow f_{ij}^{\ \ k} \overset{T_j}\longrightarrow Q_i^{\  jk} 
\overset{T_i}\longrightarrow R^{ijk}.
\nn\ee
The T-duality pattern results from canonical transformations exchanging position and momentum operators in the
 phase space.

The role of the phase space in this framework becomes even more central under the realization that certain 
non-associative structures emerge when the compactification conditions are solved. Essentially there are two 
types of such structures: (a) non-associativity of the dynamical degrees of freedom of the theory, 
i.e. the Hermitian matrices and (b) non-associativity between gauge transformations represented by unitary 
operators. They both have a common origin in the non-associativity of the phase space algebra. However, 
since the above quantities should be operators on a Hilbert space, their non-associativity has to be resolved \cite{Jackiw1,Jackiw2}.

The resolution of non-associativity in the above two cases follows a different path. In the case of the 
dynamical degrees of freedom, where non-associativity appears only in the solution associated to a non-geometric
$R$ flux, the interpretation is based on a duality between position space and momentum space. 
Thus, a solution which is not well-defined in position space turns out to be perfectly well-defined in momentum 
space. A similar argument holds for the $Q$ flux case. Thus, while the solutions related to geometric and NS-NS
fluxes are defined in position space, the ones for the non-geometric fluxes are defined on momentum space. 
What is more, in momentum space a $Q$ flux plays the role of a geometric flux (i.e. a connection in 
supergravity language), while the $R$ flux plays the role of the $H$ flux (i.e. a tensor field in 
supergravity). The latter observation indicates that much like the non-commutativity of coordinates 
exhibits a reciprocal relation to the B-field, the non-commutativity of momenta should be reciprocally 
related to the bivector $\beta$ of generalized geometry.

As for the translation operators, the requirement of associativity leads to a quantization condition for the 
flux. Therefore, fluxes in compactified Matrix theory appear to be quantized due to this requirement. 
This is a very 
welcome feature of the matrix model, since in a quantum theory charges have to be quantized anyway.

Finally, we discussed some aspects of the effective gauge theory obtained from these compactifications. 
The transformations of the gauge fields were provided and the effective action obtained by 
inserting the solutions back into the tree-level action of the matrix model was determined. It is 
notable that in the NS-NS flux case we obtained terms which are related to the Myers and  
Chern-Simons terms. An interesting next step would be to calculate the 1-loop effective action and try to 
compare it with the Dirac-Born-Infeld and Chern-Simons actions.

\vspace{20pt}

\paragraph{Acknowledgements.}

We would like to thank Ricardo Schiappa and Huyn Seok Yang for useful discussions. 
Heartfelt thanks go to Nikos Prezas 
for numerous discussions on topics related to the present work over the previous years.
This work was partially supported by the SFB-Transregio TR33
``The Dark Universe" (Deutsche Forschungsgemeinschaft),
the European Union 7th network program ``Unification in the
LHC era" (PITN-GA-2009-237920) and the Alexander von Humboldt 
foundation.
A. C. is grateful to the Simons Centre for Geometry and Physics and the organizers of the Simons Summer Workshop in Mathematics and Physics 2012 for warm hospitality and a stimulating working atmosphere.

\end{document}